\documentstyle[preprint,aps]{revtex}
\begin{document}
\title{Baryon Magnetic Moments in a Relativistic Quark Model}
\author{Simon Capstick}
\address{Supercomputer Computations Research Institute
  and Department of Physics,\\
  Florida State University,
  Tallahassee, FL 32306}
\author{B. D. Keister}
\address{
  Department of Physics,
  Carnegie Mellon University,
  Pittsburgh, PA 15213}
\date{\today}
\maketitle
\begin{abstract}
  Magnetic moments of baryons in the ground-state octet and decuplet
  are calculated in a light-front framework.  We investigate the
  effects of quark mass variation both in the current operator and in
  the wavefunctions.  A simple fit uses single oscillator
  wavefunctions for the baryons and allows the three flavors of quark
  to have nonzero anomalous magnetic moments.  We find a good fit to
  the data without allowing for strange quark contributions to the
  nucleon moments.  A slightly better fit is obtained by allowing for
  explicit SU(3)$_f$ breaking in the wavefunctions through a simple
  mechanism.  The predictions for magnetic moments in our relativistic
  model are also much less sensitive to the values chosen for the
  constituent quark masses than those of nonrelativistic models.
  Relativistic effects can be of order 20\% in general, and can alter
  familiar relationships between the moments based on SU(3)$_f$ and a
  nonrelativistic treatment of spin.
\end{abstract}
\newpage
\narrowtext
\section{Introduction}
In the simple additive quark model without sea quark degrees of
freedom, and in the nonrelativistic limit, the magnetic moments of the
baryons considered here are given by
\begin{eqnarray}
&&\mu_p = {4\over 3} \mu_u - {1\over 3} \mu_d;\quad
\mu_n = {4\over 3} \mu_d - {1\over 3} \mu_u 
\nonumber\\
&&\mu_\Lambda = \mu_s 
\nonumber\\
&&\mu_{\Sigma^+} = {4\over 3} \mu_u - {1\over 3} \mu_s ;\quad
\mu_{\Sigma^0\to \Lambda^0} = -{1\over \sqrt{3}}(\mu_u-\mu_d) ;\quad
\mu_{\Sigma^-} = {4\over 3} \mu_d - {1\over 3} \mu_s 
\nonumber\\
&&\mu_{\Xi^0} = {4\over 3} \mu_s - {1\over 3} \mu_u ;\quad
\mu_{\Xi^-} =  {4\over 3} \mu_s - {1\over 3} \mu_d 
\label{NRmus}\\
&&\mu_{\Delta^{++}} = 3 \mu_u ;\quad
\mu_{\Delta^+} = 2 \mu_u + \mu_d ;\quad
\mu_{\Delta^0} = 2 \mu_d + \mu_u ;\quad
\mu_{\Delta^-} = 3 \mu_d \nonumber\\
&&\mu_{\Sigma^{*+}} = 2 \mu_u + \mu_s  ;\quad
\mu_{\Sigma^{*0}} = \mu_u + \mu_d + \mu_s ;\quad
\mu_{\Sigma^{*-}} = 2 \mu_d + \mu_s \nonumber\\ 
&&\mu_{\Xi^{*0}} = 2 \mu_s + \mu_u ;\quad
\mu_{\Xi^{*-}} = 2 \mu_s + \mu_d \nonumber\\
&&\mu_{\Omega^-} = 3\mu_s. \nonumber
\end{eqnarray}
These equations are generalized by Karl~\cite{Karl} to include
possible contributions of a polarized strange quark sea to the
magnetic moments of the nucleons. Since his model is based on the
assumption of SU(3)$_f$ symmetry among the ground state baryons, the
presence of this polarized strange quark sea implies additional
contributions to the moments of other baryons. These generalized
Sehgal equations~\cite{Sehgal} reduce to Eqs.~(\ref{NRmus}) by using
the nonrelativistic quark model values of $\Delta u={4\over 3}$,
$\Delta d=-{1\over 3}$, and $\Delta s = 0$, where the $\Delta q$ are
the contributions of the quarks (and of antiquarks of the same flavor,
in general) of a given flavor to the axial-vector current of the
proton for which $g_A = \Delta u - \Delta d = 5/3$.  The $\Sigma^0\to
\Lambda^0$ moment is a transition magnetic moment which is extracted
from measurements of the amplitudes for the decay $\Sigma^0\to
\Lambda^0\gamma$ (see Ref.~\cite{DPB} and Appendix A). Note that the
physical $\Sigma^0$ and $\Lambda^0$ states are not pure flavor
eigenstates but are mixed by isospin-breaking interactions, and we
consider the effects of this mixing on both the $\Lambda^0$ moment and
this transition moment.  Of the decuplet moments only those of the
$\Delta^{++}$ and $\Omega^-$ are known; the $\Delta^{++}$ moment is
extracted (with some uncertainty) from $\pi^+ p$ bremsstrahlung
data~\cite{Bosshard}.

A constituent quark model represents a significant truncation of a
Lagrangian field theory like QCD, but it captures many important
degrees of freedom and permits a systematic analysis of a large body
of data.  The parameters of the model should be interpreted as those
of the original current quarks, but substantially dressed by
nonperturbative effects of QCD.  Thus, one would expect that the quark
masses and other properties such as their magnetic moments would be
substantially renormalized.

Eqs.~(\ref{NRmus}) allow the inclusion of anomalous magnetic moments
for the quarks, if the three quark moments $\mu_u$, $\mu_d$, and
$\mu_s$ are considered as free parameters.  Note that the effective
additive magnetic moment of a strange quark is assumed to be the same,
for example, in the $\Lambda^0$ state (where the other quarks are
light) and in the $\Omega^-$ state (composed entirely of strange
quarks). As pointed out by Karl, this assumption may not hold if the
sizes of the baryons are dependent on their quark structure [which
introduces explicit breaking of SU(3)$_f$]. It is also possible that
relativistic effects, both kinematical and dynamical, can modify these
relations. Our calculation allows us to explore these possibilities in
a simple way.

Previous relativistic work based on light-front
dynamics~\cite{Aznaurian,McKellar,Schlumpf_oct,CC,CPSS} has shown that
it is impossible to fit simultaneously the proton and neutron magnetic
moments without some sort of modification of the quark model
parameters.  These calculations were also carried out in the absence
of a strange quark contribution to the proton electromagnetic
currents.  They employed a Gaussian wave function for the nucleons of
the form $\exp[-M_0^2/\beta^2]$, where $M_0$ is the non-interacting
mass of the three-quark system and $\beta$ is a size
parameter~\cite{betaNorm}, which permits simple analytic calculations
but which cannot easily be extended to a complete set of orthonormal
wave functions which can be mixed via a realistic interaction.
Aznauryan, et al., considered magnetic moments and weak decay
constants in the baryon octet, varying the anomalous quark moments to
achieve a fit~\cite{Aznaurian}.  Tupper, et al.~\cite{McKellar}, Chung
and Coester~\cite{CC}, and Cardarelli, et al.~\cite{CPSS} examined the
sensitivity of the fits to variations in the quark parameters (mass,
anomalous moment, etc.). For some reasonable values of the baryon
parameters, this meant adopting anomalous moments for the light quarks
which are not proportional to their charges.  Schlumpf calculated
magnetic moments and weak decay constants in the baryon
octet~\cite{Schlumpf_oct} and decuplet~\cite{Schlumpf_dec}, and varied
the size parameter $\beta$ separately for the two subclusters in the
three-quark system, i.e., a quark-diquark picture.

Our approach is similar to several of the earlier works cited above,
except that we provide further generality to the calculations.
In particular, we study the effects of unequal quark masses, not just
in the quark magnetic moments, but in the wave functions and current
matrix elements.  The spatial wavefunctions are oscillator ground
states characterized at first by a single momentum parameter $\beta$,
but then are given two momentum parameters which depend kinematically
upon the quark masses.

We calculate magnetic moments in a light-front framework for the
entire ground state baryon octet and decuplet.  We also discuss our
results in light of some recent direct calculations of quark anomalous
magnetic moments via meson loops.
\section{Outline of Calculation}
The elements of the calculation of baryon light-front current matrix
elements are described in detail in Ref.~\cite{CK}.  We present here a
brief summary of the important features of the calculation of matrix
elements and the extraction of magnetic moments.

Free-particle state vectors 
$| {\tilde{\bf p}}\mu\rangle $ are labeled by the light-front vector 
${\tilde{\bf p}} = (p^+,{\bf p}_\perp)$ and are normalized as follows:
\begin{equation}
\langle {\tilde{\bf p}}' \mu'| {\tilde{\bf p}} \mu\rangle 
= (2\pi)^3 \delta_{\mu'\mu} \delta({\tilde{\bf p}}' - {\tilde{\bf p}}).
\label{AD}
\end{equation}
A calculation of the matrix elements
\begin{equation}
  \langle M j; {\tilde{\bf P}}'\mu' | I^+(0) | M j; {\tilde{\bf
      P}}\mu\rangle \to \langle {\tilde{\bf P}}'\mu' | I^+(0) |
  {\tilde{\bf P}}\mu\rangle
\label{MZAA}
\end{equation}
is sufficient to determine all Lorentz-invariant form factors for a
baryon of mass~$M$ and spin~$j$.  The current operator is taken to be
the sum of single-quark operators with light-front spinor matrix
elements given by
\begin{equation}
  \langle {\tilde{\bf p}}'\mu' | I^+(0) |{\tilde{\bf p}}\mu\rangle
  =F_{1q}(Q^2)\delta_{\mu' \mu}-i(\sigma_y)_{\mu' \mu}{Q\over
    2m}F_{2q}(Q^2),
\label{qlfme}
\end{equation}
where $Q\simeq \sqrt{-q^\nu q_\nu}$.  The momentum wavefunctions are
expressed as follows:
\begin{eqnarray}
  \langle {\tilde{\bf p}}_1 \mu_1 {\tilde{\bf p}}_2 \mu_2 {\tilde{\bf
      p}}_3 \mu_3| {\tilde{\bf P}}\mu \rangle &&=
  \left|{\partial({\tilde{\bf p}}_1, {\tilde{\bf p}}_2, {\tilde{\bf
          p}}_3) \over \partial({\tilde{\bf P}}, {\bf k}_1, {\bf
        k}_2)}\right| ^{-{1\over2}} (2\pi)^3 \delta({\tilde{\bf p}}_1
  + {\tilde{\bf p}}_2 + {\tilde{\bf p}}_3 - {\tilde{\bf P}}) \nonumber
  \\ &&\quad\times \langle {\textstyle{1\over2}}
  {\bar\mu}_1{\textstyle{1\over2}} {\bar\mu}_2 | s_{12}\mu_{12}\rangle
  \langle s_{12}\mu_{12} {\textstyle{1\over2}}{\bar\mu}_3 | s
  \mu_s\rangle
\langle l_\rho \mu_\rho l_\lambda \mu_\lambda | L \mu_L \rangle 
\langle L \mu_L s \mu_s | j \mu\rangle  \nonumber  \\
&&\quad\times
Y_{l_\rho\mu_\rho}({\hat{\bf k}}_\rho)
Y_{l_\lambda\mu_\lambda}({\hat{\bf K}}_\lambda)
\Phi(k_\rho, K_\lambda) \nonumber  \\
&&\quad\times
D^{({\textstyle{1\over2}}){\dag}}_{{\bar\mu}_1\mu_1}
[{\underline R}_{cf}({k}_1)]
D^{({\textstyle{1\over2}}){\dag}}_{{\bar\mu}_2\mu_2}
[{\underline R}_{cf}({k}_2)]
D^{({\textstyle{1\over2}}){\dag}}_{{\bar\mu}_3\mu_3}
[{\underline R}_{cf}({k}_3)],
\label{MZAC}
\end{eqnarray}
where the $\mu_i$ are light-front quark spin projections, ${\tilde{\bf
    p}}_i$ the light-front quark momenta,  $\Phi(k_\rho, K_\lambda)$
is the orbital momentum wavefunction, and ${\underline
  R}_{cf}({k}_3)$ is a Melosh rotation:
\begin{equation}
  \label{Melosh}
  {\underline R}_{cf}({k})
  ={ {(p^+ +m) - i\,{\mbox{\boldmath $\sigma$}}
      \cdot\hat{\bf n}\times{\bf p}_\perp} \over
    {\left[2 p^+ (p^0 +m)\right]^{1\over 2}} }.
\end{equation}
The quantum numbers of the state vectors correspond to irreducible
representations of the permutation group.  The spins $(s_{12}, s)$ can
have the values $(0, {\textstyle{1\over2}})$, $(1,
{\textstyle{1\over2}})$ 
and $(1, {3\over2})$,
corresponding to quark-spin wavefunctions with mixed symmetry
($\chi^\rho$ and $\chi^\lambda$) and total symmetry ($\chi^S$),
respectively.  The momenta
\begin{eqnarray}
  {\bf k}_\rho &&\equiv {1\over\sqrt{2}} ({\bf k}_1 - {\bf k}_2)
  \nonumber \\ {\bf K}_\lambda &&\equiv {1\over\sqrt{6}} ({\bf k}_1 +
  {\bf k}_2 - 2{\bf k}_3)
\label{MZAD}
\end{eqnarray}
preserve the appropriate symmetries under various exchanges of
${\bf k}_1$, ${\bf k}_2$ and ${\bf k}_3$, the quark three-momenta in
the baryon rest frame.

The set of state vectors formed using Eq.~(\ref{MZAC}) and Gaussian
functions of the momentum variables defined in Eq.~(\ref{MZAD}) is
complete and orthonormal.  Since they are eigenfunctions of the
overall spin, they satisfy the relevant rotational covariance
properties. For this work, the spatial wavefunctions are taken to be
oscillator ground states of the form
\begin{equation}
\Phi(k_\rho,K_\lambda)={1\over 
  \pi^{3\over 2}\beta_\rho^{3\over 2}\beta_\lambda^{3\over 2}}
e^{-\left({k_\rho^2\over 2\beta_\rho^2}
+{K_\lambda^2\over 2\beta_\lambda^2}\right)},
\end{equation}
where $\beta_\rho$ and $\beta_\lambda$ are for the moment set equal to
a single parameter for all of the states considered.  Later we
consider the effects of allowing $\beta_\rho$ and $\beta_\lambda$ to
vary by means of a simple (nonrelativistic harmonic oscillator)
formula in terms of the masses of the quarks involved. We will show in
what follows that the dependence of our results for magnetic moments
on the form of the spatial wavefunctions (through the oscillator
parameters) is weak, which partially justifies our use of these simple
wavefunctions.

The set of state vectors formed using Eq.~(\ref{MZAC})
and Gaussian functions of the momentum variables defined in
Eq.~(\ref{MZAD}) is complete and orthonormal.  Since they are
eigenfunctions of the overall spin, they satisfy the relevant
rotational covariance properties.

While the matrix elements of $I^+(0)$ are sufficient to determine all
form factors, they are in fact not independent of each other.  Parity
considerations imply that
\begin{equation}
\langle {\tilde{\bf P}}' -\mu' | I^+(0) | {\tilde{\bf P}} -\mu\rangle 
= (-1)^{\mu' - \mu}
\langle {\tilde{\bf P}}'\mu' | I^+(0) | {\tilde{\bf P}}\mu\rangle 
\label{MMA}
\end{equation}
This cuts the number of independent matrix elements in half.
For elastic scattering, time-reversal symmetry provides another
constraint: 
\begin{equation}
\langle {\tilde{\bf P}}' \mu | I^+(0) | {\tilde{\bf P}} \mu'\rangle 
= (-1)^{\mu' - \mu}
\langle {\tilde{\bf P}}'\mu' | I^+(0) | {\tilde{\bf P}}\mu\rangle 
\label{MMAA}
\end{equation}

In addition, there can be constraints which come from the requirement
of rotational covariance of the current operator.  These can be
expressed in terms of relations among the matrix elements of
$I^+(0)$~\cite{BKANG}:
\begin{equation}
\sum_{\lambda'\lambda}
D^{j{\dag}}_{\mu'\lambda'}({\underline R}'_{ch})
\langle {\tilde{\bf P}}'\lambda' | I^+(0) | {\tilde{\bf P}}\lambda\rangle 
D^j_{\lambda\mu}({\underline R}_{ch}) = 0,\quad |\mu'-\mu| \ge 2,
\label{BAA}
\end{equation}
where
\begin{equation}
{\underline R}_{ch} = {\underline R}_{cf}({\tilde{\bf P}}, M)
{\underline R}_y({\pi\over2}), \quad
{\underline R}'_{ch} = {\underline R}_{cf}({\tilde{\bf P}}', M)
{\underline R}_y({\pi\over2}).
\label{BAABA}
\end{equation}

For $J^\pi={1\over2}^+$ baryons,
there are four matrix elements of $I^+(0)$, of which only two are
independent due to parity symmetry.  Because $j={1\over2}$, there is
no nontrivial constraint due to rotational covariance.
The baryon form factors $F_1$ and $F_2$ are determined directly via
\begin{equation}
  \langle {\tilde{\bf P}}'\mu' | I^+(0) |{\tilde{\bf P}}\mu\rangle
  =F_{1}(Q^2)\delta_{\mu' \mu}-i(\sigma_y)_{\mu' \mu}{Q\over
    2M}F_{2}(Q^2).
\label{Nlfme}
\end{equation}

For $J^\pi={3\over2}^+$ baryons, there are 16 matrix elements of
$I^+(0)$, of which eight are independent due to parity symmetry.
Time-reversal symmetry eliminates two more matrix elements.  Thus,
there are six independent matrix elements of $I^+$, which can be
chosen without loss of generality to be
$\langle{+{3\over2}}|I^+(0)|{\pm{3\over2}}\rangle$,
$\langle{+{3\over2}}|I^+(0)|{\pm{1\over2}}\rangle$ and
$\langle{+{1\over2}}|I^+(0)|{\pm{1\over2}}\rangle$.  There are three
constraints due to rotational covariance, but one of these is
redundant due to time-reversal symmetry.  Thus, only four of the above
six matrix elements should be truly independent under rotational
symmetry.  For the results reported here, we choose the matrix
elements $\langle{+{1\over2}}|I^+(0)|{\pm{1\over2}}\rangle$,
$\langle{+{3\over2}}|I^+(0)|{+{1\over2}}\rangle$ and
$\langle{+{3\over2}}|I^+(0)|{+{3\over2}}\rangle$, since they
correspond to the lowest light-front spin transfer values.

Full rotational symmetry is a dynamical constraint on light-front
calculations, and can only be fully satisfied by introducing two-body
current matrix elements.  However, the constraint due to rotational
covariance is proportional to $Q^2$ as $Q^2\to 0$.  This means that it
has no effect on calculations of magnetic moments.  Nevertheless,
there can still be relativistic effects of $O(Q)$, {\it i.e.,} to
arbitrary order in $1/m$.

Extraction of the magnetic moment of $J^P={\scriptstyle 3\over 2}^+$
states proceeds by calculation of the relativistic canonical-spin
matrix elements of 
the operator $i{\bf S}_X\times {\bf q}$, where ${\bf S}_X$
is the spin of the baryon $X\in\{\Delta,\Omega\}$. The Sachs form factors
of the $X$ state are defined in terms of canonical spins by the
relation~\cite{WA} 
\begin{eqnarray}
  {}_c\langle X p^\prime s^\prime|{\bf I}(0)|X ps\rangle_c&=&
  {e\over 2M_X} \psi^{\dag}_{X,s^\prime}
  \Biggl[\left(G^X_{E0}+G^X_{E2}
    \left[{\bf S}^{[2]}\times {\bf q}^{[1]}\times {\bf q}^{[1]}\right]^{[0]}
  \right){\bf P}\nonumber \\
  &+&i\left(G^X_{M1}{\bf S}_X+G^X_{M3}
    \left[{\bf S}^{[3]}_X\times \left[{\bf q}^{[1]}\times
        {\bf q}^{[1]}\right]^{[2]}\right]^{[1]}\right)\times {\bf q}\Biggr]
  \psi_{X,s},
\end{eqnarray}
and the reduced matrix elements of the spin operators 
\begin{eqnarray}
  {}_c\langle{\textstyle{3\over2}}||{\bf
    S}_X||{\textstyle{3\over2}}\rangle_c &=&2\sqrt{15},\nonumber \\ 
  {}_c\langle{\textstyle{3\over2}}||{\bf
    S}^{[2]}_X||{\textstyle{3\over2}}\rangle_c
  &=&-\sqrt{10/3},\nonumber \\ {}_c\langle{\textstyle{3\over2}}||{\bf
    S}^{[3]}_X||{\textstyle{3\over2}} \rangle_c &=&-(7/3)\sqrt{2/3}
\end{eqnarray}
and
$q^{[2]}=[q^{[1]}\times q^{[1]}]^{[2]}=\sqrt{8\pi\over 15}{\bf q}^2
Y_{[2]}(\Omega_q)$.
For $\lambda = -{\textstyle{1\over2}}$, $\lambda' =
+{\textstyle{1\over2}}$, and momentum transfer along the $z$ axis
(conventional Breit frame), 
\begin{equation}
  {}_c\langle {\textstyle{3\over2}} +{\textstyle{1\over2}}| i[{\bf
    S}_X\times {\bf q}]_{1\mu} | {\textstyle{3\over2}}
  -{\textstyle{1\over2}}\rangle_c
  = -4\sqrt{2} Q.
\end{equation}
Thus, for the $+1$ spherical tensor component of the
current~\cite{KP},
\begin{equation}
  -{1\over\sqrt{2}}{}_c\langle {\textstyle{3\over2}}
  +{\textstyle{1\over2}}| I^1(0) + i I^2(0) | {\textstyle{3\over2}}
  -{\textstyle{1\over2}}\rangle_c = -4\sqrt{2} \,{Q\over {2 M_X}}\,
  G_{M1}.
\end{equation}
\section{Results}
As a direct measure of the size of relativistic effects in the
calculation of the baryon magnetic moments, we show in
Table~\ref{NRtoR} a comparison between magnetic moments calculated
using the nonrelativistic formulae of Eqs.~(\ref{NRmus}) and our
relativistic approach. We have restricted this calculation to baryons
for which moment data exist, using the quark masses $m_{u,d}=330$ MeV
and $m_s=550$ MeV [chosen roughly to fit the moments using the
nonrelativistic formulae in Eqs.~(\ref{NRmus})] and
$\beta_\rho=\beta_\lambda=0.41$~GeV in our relativistic calculation,
and have found the corresponding nonrelativistic moments using
Eqs.~(\ref{NRmus}). This value of harmonic oscillator parameter has
been shown roughly to fit the nucleon form factors when calculated in
a relativistic model with single-oscillator wavefunctions in
Ref.~\cite{CK}. Note that the relativistic calculation uses the
physical mass for the baryon, rather than the sum of the quark masses,
when calculating kinematical quantities.

Interactions between the quarks which distinguish between the $u$ and
$d$ quarks can cause an isospin-violating mixing between the $\Sigma^0$ and
$\Lambda^0$. As the two-state mixing angle $\theta_{\Sigma\Lambda}$ is about
+0.0135 radians~\cite{SLmixing}, the physical states $\Sigma^0$ and
$\Lambda^0$ can to a good approximation be written in terms of the SU(3)$_f$
flavor eigenstates $\bar{\Sigma}^0$ and $\bar{\Lambda}^0$ as
\begin{eqnarray}
\Sigma^0&=&\bar{\Sigma}^0 - \theta_{\Sigma\Lambda} \bar{\Lambda}^0\nonumber\\
\Lambda^0&=&\bar{\Lambda}^0 + \theta_{\Sigma\Lambda} \bar{\Sigma}^0.
\end{eqnarray}
This mixing affects the $\Lambda^0$ moment as well as the
$\Sigma^0\to\Lambda^0$ transition moment (and in principle also the
unmeasured $\Sigma^0$ moment) by
\begin{eqnarray}
\mu_{\Lambda^0}&=&\mu_{\bar{\Lambda}^0} 
+2\theta_{\Sigma\Lambda}\mu_{\bar{\Sigma}^0\to \bar{\Lambda}^0}\nonumber\\
\mu_{\Sigma^0\to\Lambda^0}&=&\mu_{\bar{\Sigma}^0\to \bar{\Lambda}^0} 
+\theta_{\Sigma\Lambda}
\left(\mu_{\bar{\Sigma}^0}-\mu_{\bar{\Lambda}^0}\right).
\end{eqnarray}
The results shown for $\mu_{\Lambda^0}$ and $\mu_{\Sigma^0\to \Lambda^0}$ in
Table~\ref{NRtoR} (and below in Table~\ref{fits}) are inclusive of the mixing
correction to the flavor eigenstate moments which we calculate in our
model. The effect is to lower $\mu_{\Lambda^0}$ by about -0.04~$\mu_N$, and
raise $\mu_{\Sigma^0\to\Lambda^0}$ by between 0.01 and 0.02~$\mu_N$.

The second column in Table~\ref{fits} shows the result of reducing the
constituent quark masses to $m_{u,d}=220$ MeV and $m_s=419$ MeV, which
are the values which fit the meson and baryon spectra~\cite{CI,GI}
and, more importantly, the mass splittings between the various charge
states~\cite{isospin} in the relativized quark model. One clear
advantage of our relativistic calculation is that the resulting
magnetic moments are quite insensitive to the quark masses. For
example, we see by comparing Table~\ref{NRtoR} with the second column
in Table~\ref{fits} that the proton magnetic moment changes from 2.42
$\mu_N$ to 2.76 $\mu_N$ (a 14\% enhancement) when the inverse quark
mass is raised by 50\%.

It is useful to discuss our results in light of the work of Chung and
Coester~\cite{CC}, who fit their relativistic calculation of the moments to a
linear function of $\beta/m_q$, with $m_q$ the light quark mass, with
the result (in units of nuclear magnetons):
\begin{eqnarray}
  \mu_p-1&\simeq& {M_N\over m_q}\left\{ {2\over 3}
    \left(1-0.19{\beta\over m_q}\right) +\left[{4\over
        3}F_{2u}(0)-{1\over 3}F_{2d}(0)\right]
    \left(1-0.097{\beta\over m_q}\right)\right\}\nonumber\\ 
  \mu_n&\simeq&{M_N\over m_q}\left\{ -{2\over
      3}\left(1-0.225{\beta\over
        m_q}\right)+\left[{4\over 3}F_{2d}(0)-{1\over 3}F_{2u}(0)\right]
        \left(1-0.097{\beta\over m_q}\right) \right\},
\label{CCmuN}
\end{eqnarray}
where $F_{2q}$ is the anomalous magnetic moment of the quark. Note
that these formulae agree with the relations in Eqs.~(\ref{NRmus}) in
the limit $\beta/m_q\to 0$ and when $M_N=3m_q$ if we take
$\mu_q=[e_q+F_{2q}(0)]e/(2m_q)$, where $e_u=+2/3$, $e_d=-1/3$, etc. In
these formulae the terms proportional to $\beta/m_q$ are corrections
from relativity to the contributions of the Dirac and Pauli moments of
the quarks.

Relativistic effects tend to reduce the baryon magnetic moments, the
primary effect coming from Melosh rotations of the quarks.  This fact
has been known for some time~\cite{Close}, and is reflected in factors
like $(1-k\beta/m_q)$ in Eq.~\ref{CCmuN}.  Lowering the quark mass
raises the quark Dirac moment $e_q/2m_q$ but lowers this factor, with
the result that the moment of the baryon is largely unaffected, as
seen in our calculated results.

In Ref.~\cite{CPSS} essentially the same procedure was adopted, but
with more sophisticated configuration-mixed wavefunctions resulting
from a global fit to the spectrum~\cite{CI}. These wavefunctions tend
to have larger relativistic effects and so the nucleon anomalous
moments, as we can see from Eq.~(\ref{CCmuN}), are reduced in
magnitude, with the reduction in the neutron moment being larger than
that in the proton moment. This is offset by adopting an
anomalous magnetic moment for the down quark of
$F_{2d}(0)=:\kappa_d=-0.153$ and a smaller moment for the up quark of
$F_{2u}(0)=:\kappa_u=0.085$.  Note that these are substantial
anomalous moments, yielding quark moments of $\mu_u=1.13(+{2\over
  3}){e\over 2m_q}$ and $\mu_d=1.46(-{1\over 3}){e\over 2m_q}$.

When we adopt baryon wavefunctions which all have the same spatial
size the SU(6) symmetry represented by the relations from
Eqs.~(\ref{NRmus}) for the octet baryon magnetic moments is broken,
but those relations partially apply to our relativistic decuplet
baryon results. This is because the spin and spatial wavefunctions of
the decuplet baryons have separate total permutational symmetry.  One
can think of each quark as providing an effective additive moment, as
defined by Eqs.~(\ref{NRmus}).  The effective additive moment of a
quark depends on its environment, which in this case means the masses
of the other quarks in the baryon.  Fitting Eqs.~(\ref{NRmus}) to our
relativistic results in the second column of Table~\ref{fits} yields
the effective additive quark moments $\mu_u=1.82$, $\mu_d=-0.91$ in
the $\Delta$ states, $\mu_u=1.77$, $\mu_d=-0.89$, and $\mu_s=-0.63$ in
the $\Sigma^*$ states, $\mu_u=1.75$, $\mu_d=-0.88$, and $\mu_s=-0.62$
in the $\Xi^*$ states, and $\mu_s=-0.61$ in the $\Omega^-$. Note that
in all cases the effective additive moments of the light quarks are in
the ratio of their charges. There is a slight dilution of the
effective moments of the quarks when in the environment of heavier
quarks.

Our results appear to be consistent with those of McKellar {\it et
al.}~\cite{McKellar}, who choose not to adopt anomalous moments for
the quarks, but instead concentrate on examining the dependence of the
moments on the masses of the light and strange quarks and the
oscillator parameter $\beta$. They conclude that there is no choice of
$\beta$ and light-quark mass for which the nucleon moments are
reproduced, and that a fit to the octet data is not possible without the
inclusion of quark anomalous moments.

In the third column of Table~\ref{fits} we have shown the result of
fitting the nine precisely determined magnetic moments by allowing
nonzero anomalous magnetic moments of the quarks (we did not allow the
masses to vary from the values prescribed by
Refs.~\cite{CI,GI,isospin}, or vary the harmonic oscillator constant
from the value 2.08 fm$^{-1}$).  This three-parameter fit reduces the
root-mean-square deviation of the calculated moments from the nine
precise data from 0.135~$\mu_N$ to 0.076~$\mu_N$.  The resulting quark
anomalous moments are rather small: $\kappa_u=-0.011$,
$\kappa_d=-0.048$, and $\kappa_s=-0.020$ in units of quark magnetons
[for each flavor of quark $\kappa_q=F_{2q}(0)$, so that $\kappa_q$ is
in units of $\mu_q=e/2m_q$, where $e$ is the electron charge, see
Eq.~(\ref{qlfme})].  Equivalently, if we define a quark moment by
$\mu_q=[F_{1q}(0)+F_{2q}(0)]{e\over 2m_q}$, we have $\mu_u=0.98
(+{2\over 3}){e\over m_u}$, $\mu_d=1.14(-{1\over 3}){e\over m_d}$, and
$\mu_s=1.06(-{1\over 3}){e\over m_s}$.

In the fourth column of Table~\ref{fits}, we have shown the
effects of adopting a simple nonrelativistic dependence of the
harmonic oscillator size parameters on the masses of the quarks in the
baryon, {\it i.e.}
\begin{equation}
\beta_\rho=(3Km)^{1\over 4},\ \ \beta_\lambda=(3Km_\lambda)^{1\over
  4},
\label{NRalphas}
\end{equation}
where $m_\rho=m_1=m_2$ is the mass of the two equal mass quarks and 
\begin{equation}
m_\lambda={3mm_3\over 2m+m_3},
\end{equation}
where $m_3$ is the mass of the odd quark out.  For strangeness-zero
baryons we have used $\beta_\rho=\beta_\lambda=0.410$~GeV when
$m=m_3=0.220$~GeV, so solving for $3K$ ($K$ is the oscillator
constant) we find $3K=0.128$~GeV$^3$. Note that this differs from the
approach taken by Schlumpf~\cite{Schlumpf_oct,Schlumpf_dec} who
concentrated on calculating weak decay constants in terms of strongly
asymmetric $\beta$ parameters, which imply significant diquark
clustering in the baryon wavefunctions. Such strong clustering in the
wavefunctions does not arise in the relativized quark model of the
spectrum of these states~\cite{CI}.

As can be seen from
Table~\ref{fits}, our results are largely insensitive to changes in
the wavefunction, here made through the simple mechanism of changing
the harmonic oscillator scale. This can be easily understood, as this
scale only enters in the relativistic corrections to the
formulae~(\ref{NRmus}). Equivalently, details of the wavefunction can
only affect the size of the change in the moments due to relativistic
corrections illustrated in Table~\ref{NRtoR}. However, a slightly
better fit to the moments is obtained when using these wavefunctions,
with the result that the r.m.s.\ deviation from the data for the
nine precisely measured moments is reduced to 0.068 $\mu_N$. The quark
anomalous moments that give this fit are essentially unchanged at
$\kappa_u=-0.006$, $\kappa_d=-0.047$ and $\kappa_s=-0.022$.

Aznauryan {\it et al.}~\cite{Aznaurian} performed a fit to the 1984
data for the moments of the octet baryons using anomalous moments for
the quarks, and make predictions for weak decay constants. They adopt
quark masses of $m_{u,d}=271$ MeV and $m_s=397$ MeV, and allow the
parameter $\beta$ in the Gaussian wavefunction $\exp[-M_0^2/\beta^2]$
to vary linearly with the average mass of the three quarks. Note that
this choice of wavefunction does not allow $\beta_\rho$ and
$\beta_\lambda$ to differ, as is required by the solution of the
dynamical problem. Their fit to the octet moments is of similar
quality to ours, with the exception of the $\Sigma^0\to \Lambda^0$ transition
moment, which is too small. The quark anomalous moments they find for
their fit are $\kappa_u=0.012$, $\kappa_d=-0.059$, and
$\kappa_s=-0.016$, similar to those found here.

Our results for the decuplet baryon moments differ from those of
Schlumpf~\cite{Schlumpf_dec} due to our adoption of different quark
masses. Schlumpf uses $m_u=m_d=260$ MeV in his relativistic fits,
which is significantly {\it larger} than our 220 MeV, and an $m_s$ of
380 MeV which is {\it smaller} than our value of 419 MeV. The same is
true of the quark masses used by Aznauryan {\it et al.} Our quark
masses are motivated by relativized fits to the meson and baryon
spectra~\cite{CI,GI}, and to isospin violations in ground state meson
and baryon masses.  A difference between the light and strange quark
masses of 120 MeV may not be large enough to be consistent with these
other constraints.

The use of quark anomalous magnetic moments is intended to account for
the fact that constituent degrees of freedom are effective, and that
such quarks receive substantial QCD dressing.  From the point of view
of chiral symmetry, one could express such quark dressing in terms of
pion loops.  Several groups have investigated this
possibility~\cite{CohenWeber,LiLiao,Ito}.  In general, the anomalous
moments which result from pion loops are significantly higher than
those obtained in our phenomenological fit.  For example, recent
results by Ito~\cite{Ito} give the ranges $\kappa_u$ = 0.0550--0.1118
and $\kappa_d$ = -(0.0832--0.1438).  His $\kappa_u$ has the opposite
sign from our phenomenological fit, and his $\kappa_d$ is several
times our value.  Using his values of $\kappa_u$ and $\kappa_d$ to
compute baryon moments would give a worse fit than one with no
anomalous moments.  On the other hand, Cohen and Weber find large
cancellations of pion loop contributions with other corrections to
baryon magnetic moments~\cite{CohenWeber}.  The lesson from this is
that a constituent quark model is not easily corrected by adding pion
loops to describe QCD quark dressing.

The physics of meson cloud effects in baryons was also studied from a
general perspective of chiral symmetry by Cheng and
Li~\cite{ChengLi95,ChengLi96}.  They characterize effects of the quark
sea in terms of parameters $g_8$ and $g_0$, corresponding to the
contributions of SU(3)$_f$ pseudoscalar octet and singlet Goldstone
bosons, respectively.  They find a reasonable fit~\cite{ChengLi95} to
the measure $\Delta\Sigma$ of the quark contribution to the proton
spin appearing in the Bjorken~\cite{Bjorken} and
Ellis-Jaffe~\cite{EllisJaffe} sum rules, as well as to the magnetic
moments of the baryon octet~\cite{ChengLi96}.

Such effects of the quark sea should also be considered in a
relativistic quark model.  Technically, the procedure will be much
more difficult since, as we have seen, the quark moments do not enter
in a simple additive fashion.  Ma and Zhang find substantial
reductions of the proton spin matrix elements, related to the axial
coupling $g_A$, which enter the Bjorken~\cite{Bjorken} and
Ellis-Jaffe~\cite{EllisJaffe} sum rules~\cite{MaZhang,Ma}.  This
property has also been noted by other authors~\cite{BrodskySchlumpf}.
Thus, the combined effects of relativity and the quark sea must be
considered together.  In that regard, it may be better to use a
variation of the approach of Cheng and Li to parameterize the effect
of the quark sea, rather than to compute specific meson loop
contributions. 
\section{Conclusions}
We have seen that it is possible to achieve a quite satisfactory fit
to the measured ground state baryon magnetic moments with the
inclusion of relativistic effects and allowing the constituent quarks
to have nonzero anomalous moments.  This seems reasonable given that
the constituent quark is an effective degree of freedom, much like the
nucleon when it is bound into a nucleus.  A measure of the
effectiveness of our model is to compare to a fit using the simple
additive model of Eqs.~(\ref{NRmus}) and three arbitrary quark
moments. The result of doing this is shown in Table~\ref{fits}, with a
root-mean-square deviation for the nine data of 0.100 $\mu_N$.
Clearly our relativistic fits improve on this. The differences are
caused by the fact that, in a relativistic model, the moments of the
quarks do not simply add to the moments of the baryons, as has been
pointed out by many other
authors~\cite{Aznaurian,McKellar,Schlumpf_oct,CC,CPSS,Schlumpf_dec}.

The anomalous moments obtained in our fit can be thought of as quark
sea effects which dress the effective degrees of freedom in such
models.  However, quark sea effects are not interchangeable with pion
loop effects, as the actual numbers obtained in our fit differ
considerably from a direct calculation of baryon moment modifications
due to pion loops.

Our results are comparable to the fit to the moments achieved by
Karl~\cite{Karl} in the presence of a strange quark contribution to
the magnetic moment of the nucleons. The root-mean-square deviation for his
fit to the eight octet moments is 0.084 $\mu_N$. We would conclude
that it is possible to improve on a simple nonrelativistic fit to the
data without strange quark contributions to the magnetic moment of the
nucleons.

The results presented here can be expected to change with the adoption
of mixed wavefunctions such as those of Ref.~\cite{CI} used by
Cardarelli {\it et al.},~\cite{CPSS}. We have made some exploratory
calculations of this type for the baryons for which data for their
moments exist, and have found that it is impossible to achieve a fit
of the quality shown in Table~\ref{fits} with the relativized model
wavefunctions of Ref.~\cite{CI}.  On the other hand, the results of
Ref.~\cite{CPSS} also reveal that that these wavefunctions have large
amplitudes at higher momentum, which may lead to an overprediction of
relativistic effects.  In Ref.~\cite{CPSS}, this behavior was
offset by giving the quarks quite soft momentum-dependent form
factors.  Another possibility is to consider quark wavefunctions which
have smaller amplitudes at higher momentum.  This question is
presently under investigation.
\section{Acknowlegements}
This work was supported in part by U.S. National Science Foundation
under Grant No.\ PHY-9319641 (BDK), by the U.S. Department of Energy
under Contract No.\ DE-AC05-84ER40150 and by the Florida State
University Supercomputer Computations Research Institute which is
partially funded by the Department of Energy through Contract
DE-FC05-85ER250000 (SC).

\section{Appendix A: The $\Sigma^0$ to $\Lambda^0$ transition moment}

In Ref.~\cite{DPB} the transition moment for $\Sigma^0$ to $\Lambda^0$
is defined to be
\begin{equation}
  \left[\mu_{\Sigma \Lambda} \over \mu_N\right]^2 = {1\over \tau} {8
    \hbar M_p^2 M_\Sigma^3 \over \alpha (M_\Sigma^2-M_\Lambda^2)^3},
\end{equation}
where $\tau$ is the lifetime of the $\Sigma^0$ (which goes almost 100\%
through $\Lambda^0$ $\gamma$). In the Particle Data Group~\cite{PDG} the
photon width of a resonance $R$ decaying to a nucleon is given by
\begin{equation}
  \Gamma_\gamma={k^2 \over \pi}{2M_N\over (2J+1)M_R}\left[\vert
    A_{1\over 2}\vert ^2+\vert A_{3\over 2}\vert^2\right],
\label{photonwidth}
\end{equation}
where $J$ is the spin of the decaying resonance and $k$ is the photon
c.m. decay momentum. Adapting this to the $\Sigma^0$ to $\Lambda^0$
decay, we have
\begin{equation}
  \Gamma={k^2\over \pi}{M_\Lambda\over M_\Sigma}\vert A_{1\over
    2}\vert ^2,
\end{equation}
Replacing the lifetime in the first equation by $\hbar\over \Gamma$,
we have
\begin{equation}
  \left[\mu_{\Sigma \Lambda} \over \mu_N\right]^2={M_p^2\Gamma\over
    \alpha k^3},
\end{equation}
where $k=(M_\Sigma^2-M_\Lambda^2)/2M_\Sigma$ is the c.m. frame decay
momentum.

The result is that we can write the transition moment directly in
terms of the helicity amplitude (calculated in the rest frame of the
decaying $\Sigma^0$) as
\begin{equation}
  \left[\mu_{\Sigma \Lambda} \over
    \mu_N\right]^2={2M_p^2M_\Lambda\over \pi\alpha
    (M_\Sigma^2-M_\Lambda^2)}\vert A_{1\over 2}\vert ^2.
\end{equation}
\pagebreak
\begin{table}
  \caption{Relativistic effects in baryon magnetic moments for which
    data exist; here $m_{u,d}=330$ MeV, $m_s=550$ MeV, and
    $\beta_\rho=\beta_\lambda=2.08$ fm$^{-1}$. Moments are in units
    of nuclear magnetons $\mu_N=e/2m_N$. Data are from
    Ref.~\protect\cite{PDG}.}
  \label{NRtoR}
  \begin{tabular}{rrrr}
    \multicolumn{1}{c} {moment} &
    \multicolumn{1}{c} {nonrel.} &
    \multicolumn{1}{c} {rel.} & 
    \multicolumn{1}{c} {data} \\ \hline \hline
    $\mu_p$             &  2.85 &  2.42 &  2.79 \\ \hline
    $\mu_n$             & -1.89 & -1.34 & -1.91 \\ \hline
    $\mu_\Lambda$       & -0.61 & -0.50 & -0.61 \\ \hline
    $\mu_{\Sigma^+}$    &  2.72 &  2.22 &  2.46 \\ \hline
    $\mu_{\Sigma^-}$    & -1.07 & -1.00 & -1.16 \\ \hline
    $\mu_{\Sigma^0 \to \Lambda^0}$   & -1.62 & -1.20 & -1.61 \\ \hline
    $\mu_{\Xi^0}$       & -1.39 & -1.09 & -1.25 \\ \hline
    $\mu_{\Xi^-}$       & -0.44 & -0.63 & -0.65 \\ \hline
    $\mu_{\Delta^{++}}$ &  5.69 &  5.05 & 3.5-7.5  \\ \hline
    $\mu_\Omega$        & -1.71 & -1.77 & -2.02 \\
  \end{tabular}
\end{table}

\begin{table}
  \caption{Baryon magnetic moments calculated in the relativistic
    model with $m_{u,d}=220$ MeV and $m_s=419$ MeV. In the second and
    third columns $\beta_\rho$ and $\beta_\lambda$ are fixed at 2.08
    fm$^{-1}$, while in the fourth column they vary according to the
    simple formula of Eq.~(\protect\ref{NRalphas}). The third and
    fourth columns are independent fits to the moments using the three
    quark anomalous moments, as described in the text. A
    nonrelativistic fit using three quark moments and
    Eqs.~(\protect\ref{NRmus}) is shown in the first column for
    comparison purposes. Moments are in units of nuclear magnetons
    $\mu_N=e/2m_N$. }
  \label{fits}
  \begin{tabular}{rccccr}
    \multicolumn{1}{c} {} &
    \multicolumn{1}{c} {} & 
    \multicolumn{1}{c} {$\kappa_i=0$} &
    \multicolumn{1}{c} {$\kappa_i$ fit} & 
    \multicolumn{1}{c} {$\kappa_i$ fit} & 
    \multicolumn{1}{c} {} \\
    \multicolumn{1}{r} {moment} &
    \multicolumn{1}{c} {NRQM fit} & 
    \multicolumn{1}{c} {$\beta_\rho=\beta_\lambda$} &
    \multicolumn{1}{c} {$\beta_\rho=\beta_\lambda$} & 
    \multicolumn{1}{c} {$\beta_\rho$, $\beta_\lambda$ from 
      Eq.~(\protect\ref{NRalphas})} & 
    \multicolumn{1}{c} {data} \\ \hline \hline
    $\mu_p$             &  2.66 &  2.76 &  2.76 &  2.78 &  2.79 \\ \hline
    $\mu_n$             & -1.94 & -1.62 & -1.82 & -1.82 & -1.91 \\ \hline
    $\mu_\Lambda$       & -0.69 & -0.61 & -0.65 & -0.63 & -0.61 \\ \hline
    $\mu_{\Sigma^+}$    &  2.54 &  2.56 &  2.52 &  2.51 &  2.46 \\ \hline
    $\mu_{\Sigma^-}$    & -1.14 & -1.08 & -1.29 & -1.29 & -1.16 \\ \hline
    $\mu_{\Sigma^0 \to \Lambda^0}$   & -1.58 & -1.45 & -1.54 & -1.52 
& -1.61 \\ \hline
    $\mu_{\Xi^0}$       & -1.45 & -1.31 & -1.35 & -1.32 & -1.25 \\ \hline
    $\mu_{\Xi^-}$       & -0.53 & -0.63 & -0.63 & -0.64 & -0.65 \\ \hline
    $\mu_{\Delta^{++}}$ &       &  5.45 &  5.33 &  5.38 & 3.5-7.5 \\ \hline
    $\mu_{\Delta^{+}}$  &       &  2.72 &  2.47 &  2.51 & \\ \hline
    $\mu_{\Delta^{0}}$  &       &  0.01 & -0.36 & -0.34 & \\ \hline
    $\mu_{\Delta^{-}}$  &       & -2.72 & -3.22 & -3.21 & \\ \hline
    $\mu_{\Sigma^{*+}}$ &       &  2.91 &  2.79 &  2.84 & \\ \hline
    $\mu_{\Sigma^{*0}}$ &       &  0.25 &  0.01 &  0.03 & \\ \hline
    $\mu_{\Sigma^{*-}}$ &       & -2.41 & -2.77 & -2.78 & \\ \hline
    $\mu_{\Xi^{*0}}$    &       &  0.51 &  0.40 &  0.42 & \\ \hline
    $\mu_{\Xi^{*-}}$    &       & -2.12 & -2.36 & -2.38 & \\ \hline
    $\mu_\Omega$        & -1.95 & -1.84 & -1.96 & -1.99 & -2.02 \\
    \hline\hline 
    $\sqrt{\sum(\mu_{\rm th}-\mu)^2/9}$ & 0.100 & 0.135 & 0.076 & 0.068\\
  \end{tabular}
\end{table}


\begin{references}
\bibitem{betaNorm} Note that the definition of the size parameter
  $\alpha$ in the single-Gaussian wavefunction in Ref.~\cite{CC} can
  only be connected to that adopted here and elsewhere in the
  literature in the nonrelativistic limit, where
  $\alpha \leftrightarrow \sqrt{3} \beta$.
\end{references}
\end{document}